\documentclass[letterpaper, 10 pt, conference]{ieeeconf}
\IEEEoverridecommandlockouts                              

\usepackage{amsmath,amsfonts}
\usepackage{algorithmic}
\usepackage{array}
\usepackage{cite}
\usepackage{tikz}
    \usetikzlibrary{calc, arrows, shapes}
    \usetikzlibrary{decorations.pathreplacing,calligraphy, positioning,backgrounds,fit,arrows.meta}
\usepackage[utf8]{inputenc}
\usepackage[T1]{fontenc}
\usepackage{graphics} 
\def\BibTeX{{\rm B\kern-.05em{\sc i\kern-.025em b}\kern-.08em
    T\kern-.1667em\lower.7ex\hbox{E}\kern-.125emX}}
\usepackage{epsfig} 
\usepackage{mathptmx} 
\usepackage{amssymb}  
\tikzset{%
  block/.style    = {draw, thick, rectangle, minimum height = 3em,
    minimum width = 3em},
  sum/.style      = {draw, circle,minimum size=1cm}, 
  input/.style    = {coordinate}, 
  output/.style   = {coordinate}, 
  gain/.style = {
  	draw, 
    isosceles triangle,
    isosceles triangle apex angle=50,
    minimum height = 3.0em,
    outer sep=0},
}
\usepackage{mathtools}
\newcolumntype{Y}{>{\centering\arraybackslash}X}

\newtheorem{thm}{Theorem}
\newtheorem{lem}{Lemma}

\newtheorem{asmp}{Assumption}
\newtheorem{defn}{Definition}

\title{\Large\bf %
Logarithmic Barrier Functions for Practically Safe Extremum Seeking Control
}
\author{Qixu Wang$^{1}$, Patrick McNamee$^{1}$, and Zahra Nili Ahmadabadi$^{1}$
\thanks{Research was sponsored by the Army Research Office under Grant
Number W911NF-24-1-0386 and the Department of the Navy, Office of Naval Research
under ONR award number N000142412269. The views, findings, conclusions, or recommendations contained in this document are those of the authors and should not be interpreted as representing the official policies or views, either expressed or implied, of the Army Research Office, the Office of Naval Research, or the U.S. Government. The U.S. Government is authorized to reproduce and distribute reprints for Government purposes notwithstanding any copyright notation herein.}
\thanks{$^{1}$Q. Wang, P. McNamee, and Z. N. Ahmadabadi are with the Department of Mechanical Engineering, San Diego State University, San Diego, CA, USA
        {\tt\small $\lbrace$qwang0429, pmcnamee5123,zniliahmadabadi$\rbrace$@sdsu.edu}}%
}

\begin{document}
\maketitle

\begin{abstract}
This paper presents a methodology for Practically Safe Extremum Seeking (PSfES), designed to optimize unknown objective functions while strictly enforcing safety constraints via a Logarithmic Barrier Function (LBF). Unlike traditional safety-filtered approaches that may induce chattering, the proposed method augments the cost function with an LBF, creating a repulsive potential that penalizes proximity to the safety boundary. We employ averaging theory to analyze the closed-loop dynamics. A key contribution of this work is the rigorous proof of practical safety for the original system. We establish that the system trajectories remain confined within a safety margin, ensuring forward invariance of the safe set for a sufficiently fast dither signal. Furthermore, our stability analysis shows that the model-free ESC achieves local practical convergence to the modified minimizer strictly within the safe set, through the sequential tuning of small parameters. The theoretical results are validated through numerical simulations.
\end{abstract}

\section{Introduction}
The pursuit of optimal performance in autonomous systems operating in uncertain environments is a central challenge in modern control theory. Extremum Seeking Control (ESC) has emerged as a useful technique for such real-time optimization tasks, particularly where the underlying model or the objective function is unknown or difficult to characterize precisely \cite{ref:krstic-2000, ref:scheinker-2024}. Fundamentally, ESC uses dither signals to perturb the system inputs, enabling the local estimation of the performance gradient. By correlating the resulting output response with these signals, the gradient of the unknown performance function $J(\theta)$, where $\theta \in \mathbb{R}^n$ denotes the system parameter vector or state, is estimated, moving the seeker towards an extremum \cite{ref:manzie-2009}. Due to its model-free nature, ESC is attractive for a wide array of applications in the source-seeking fields~\cite{ref:suttner-2023,ref:wang-2023}.

However, traditional ESC schemes primarily focus on convergence to an extremum without explicit mechanisms for constraint enforcement during the transient phase. As autonomous and adaptive systems are increasingly deployed in complex, safety-critical environments—such as robotic navigation and chemical process operations—the strict satisfaction of operational constraints becomes important. Violating these constraints can lead to system failure, environmental damage, or risks to human safety. To address this challenge, Control Barrier Functions (CBFs) have gained popularity as a systematic approach to enforce forward invariance of safe sets for nonlinear systems \cite{ref:ames-2014,ref:ames-2017,ref:panja-2024}. Recent advancements have also extended CBFs to learning-based and robust control frameworks \cite{ref:abel-2024,ref:kim-2025}. Within extremum seeking, recent literature has developed ``safety-filtered'' ESC architectures \cite{ref:williams-2022,ref:williams-2025,ref:williams-2026}. These methods correct the nominal ESC search direction by solving quadratic programs (QPs) based on estimated CBFs. However, the continuous reliance on online QP solvers introduces substantial computational overhead, and the discrete activation of safety constraints frequently induces chattering near the boundary. Additionally, the CBF design is inherently tied to the system dynamics, so porting the controller to a different platform necessitates a complete redesign. To overcome this, \cite{ref:scheinker-2017} incorporates the noise-corrupted value in the cost function for a unicycle model. However, this approach lacks formal safety guarantees, as the ESC trajectory may enter the unsafe set. 

A possible alternative to CBFs is the Logarithmic Barrier Functions (LBFs). LBFs are a classical tool in constrained optimization \cite{ref:hauser-2006} and Model Predictive Control \cite{ref:feller-2015}. References~\cite{ref:labar-2019} and \cite{ref:guay-2015} have both incorporated LBFs in ESCs but several gaps remain. Reference~\cite{ref:guay-2015} proposed a method that augments the cost function with a barrier function for the ESC, in addition to a predictor model, an auxiliary covariance, and a Lipschitz projection operator to keep parameter estimates within a known bounded set. The estimation component introduces significant implementation complexity for dynamic models. In contrast, \cite{ref:labar-2019} proposed a Lie bracket ESC method with the LBF term. However, the method in \cite{ref:labar-2019} is limited by \cite[Asmp~9]{ref:labar-2019}, which requires the constraint function (i.e.,~$-h$) to be convex. Additionally, the method does not enforce the strict safety constraint for all time, potentially allowing the system state to enter the unsafe region during transients. In this paper, we address the high complexity of \cite{ref:guay-2015} as well as the overly restrictive assumptions of \cite{ref:labar-2019}. The proposed design in this work is simpler than the design in \cite{ref:guay-2015} (requiring fewer parameters to tune) and is more applicable than \cite{ref:labar-2019} by weaker assumptions (allowing it to be deployed on a wider range of applications).

The stability analysis of the proposed method relies on foundational averaging techniques established for general nonlinear ESC \cite{ref:krstic-2000, ref:khalil-2002,ref:mcnamee-2025}. While the stability of averaged systems is typically well-understood, proving safety for the original perturbed system presents unique challenges due to the oscillatory tracking error inherent in ESC. The analysis in this paper shows that, with sequential tuning of the barrier weight $\mu$, the dither amplitude $a$, and the dither rate $\omega$, the LBF guarantees that the system trajectories remain strictly within the safe set.

\section{Problem Statement}
\label{sec:problem-statement}
This work investigates a Practically Safe Extremum Seeking (PSfES) architecture that embeds a LBF term directly into the optimization objective. We consider a static map optimization problem. The plant is described by a state $\theta \in \mathbb{R}^n$. The goal is to minimize a cost function $J(\theta)$ and satisfy the constraint $h(\theta) > 0$. In order to achieve this, we introduce a modified cost function 
\begin{equation} 
\label{eq:modified-cost-function}
    \hat{J}(\theta) = J(\theta) - \mu \log \left( h(\theta) \right)
\end{equation}
where $h(\theta): \mathbb{R}^n \to \mathbb{R}$ is a continuously differentiable function, and $\mu > 0$ is a design parameter weighting the barrier. We define the safe set $\mathcal{S} = \{ \theta \in \mathbb{R}^n \mid h(\theta) > 0 \}$. As $\theta$ approaches the boundary of $\mathcal{S}$, the condition $h(\theta) \to 0^+$ implies ${\hat{J}(\theta) \to \infty}$. Consequently, the LBF term penalizes proximity to the boundary.

To optimize the system in multiple dimensions \cite{ref:mcnamee-2025}, we define the multivariable dither signal $s : \mathbb{R} \to \mathbb{R}^n$ and the demodulation signal $m : \mathbb{R} \to \mathbb{R}^n$. The elements of these vectors are defined using distinct frequencies to ensure the gradient estimates decouple correctly in the average system. Specifically, the $i$-th elements are defined as
\begin{equation}
    s_i(\tau) = r_i \sin(\omega'_i \tau), \quad 
    m_i(\tau, a) = \frac{2}{a\, r_i} \sin(\omega'_i \tau)
\end{equation}
where $\tau = \omega t$ denotes the fast time scale common in the previous literature~\cite{ref:mcnamee-2025}, and $\omega'_i$ and $r_i$ control the relative dither rates and amplitudes, respectively. For this work, it is assumed that the $\omega'_i$'s are all rational numbers so that the signals $s(\cdot)$ and $m(\cdot, a)$ are coperiodic with a common period $T$. Following \cite{ref:mcnamee-2025}, the relative dither rates must satisfy $\omega'_i \neq \omega'_j$ and $\omega'_i + \omega'_j \neq \omega'_k$ for all distinct indices $i, j, k$. This condition ensures that the perturbation and demodulation signals satisfy the orthogonality property
\begin{equation}
    \frac{1}{T} \int_0^{T} s_i(\tau)\  m_j(\tau, a)\, d\tau = \frac{1}{a}\delta_{ij},
\end{equation}
where $\delta_{ij}$ is the Kronecker delta. The gradient-based ESC with a LBF is
\begin{equation}
    \label{eq:dynamic-full}
    \dot{\hat{\theta}} = -k \hat{g}(\omega t, \hat{\theta}, a)
\end{equation}
where $\hat{\theta}$ is the estimate of the optimal parameter, $k > 0$ is the adaptation gain, and $\hat{g}$ is the estimate of the gradient defined as 
\begin{equation}
    \hat{g}(\omega t, \hat{\theta}, a) = \hat{J}(\hat{\theta} + a s(\omega t))\ m(\omega t, a)
\end{equation}

\begin{figure}[t]
    \vspace*{0.3cm}
    \centering
    \begin{tikzpicture}[>=Latex, scale=0.82, transform shape]
    
    \node at (3.5, 0)[text width=22ex,align=center,block] (J_theta) {$J(\theta) -\mu \log(h(\theta))$};
    
    \begin{scope}[shift={(6.0,-2)}, scale=1, rotate=0]
    \node[circle, minimum size=3ex, draw=black, inner sep=0pt, outer sep=0pt] (time_j) {};
    \draw (time_j.north west) -- (time_j.south east);
    \draw (time_j.south west) -- (time_j.north east);
    \end{scope}
    
    \node at (3.5, -2) (gain_k)[regular polygon, regular polygon sides=3,
    draw, inner sep=0.5mm, outer sep=0mm,
    shape border rotate=90] {$\frac{-k}{s}$};
    
    \draw (1.1,-2) node[circle, minimum size=3ex, draw=black, inner sep=0pt, outer sep=0pt] (sum_end) {};
    \node [input,below of =sum_end] (input_sum_end) {};
    \draw (sum_end.south) -- (sum_end.north);
    \draw (sum_end.west) -- (sum_end.east);
    
    \draw[->] (J_theta.east) -- node [midway,above] {$\hat{J}$} ++(3,0) ;
    
    \draw[->] (J_theta.east) -|++(1.5,0) |- node [midway,above] {}(time_j.east);
    \draw[->] (time_j.west) -- node [midway,above] {}(gain_k.east);
    
    \draw[<-] (time_j.south) -- node [midway,right] {$m(\tau, a)$} ++(0,-1.1);
    \draw[->] (gain_k.west) -- node [midway, above] {$\hat{\theta}$} (sum_end.east);

    \draw[<-] (sum_end.south) -- node [midway,right] {$s (\tau)$} ++(0,-1.1) ;
    
    \draw[->] (sum_end.west) -| ++ (-0.8, 0) |- node [midway, above, xshift = 8.3mm ] {$\theta$} ([yshift=0mm]J_theta.west);
\end{tikzpicture}
    \caption{Block diagram of the multi-dimensional combined ESC and LBF.}
    \label{fig:flowchart-lbf}
\end{figure}
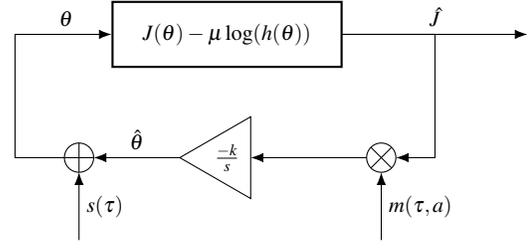
The full block diagram is shown in Fig.~\ref{fig:flowchart-lbf}. We impose the following structural assumptions on the unknown maps:
\begin{asmp}[Nominal Cost Function]      
    \label{asmp:cost-function}
    The objective function $J(\theta)$ is unknown to the controller but is assumed to be convex and quadratic:
\begin{equation} 
    \label{eq:def:cost-function-original}
    J(\theta)=J^*+\frac{1}{2} (\theta-\theta^*)^{\top} H (\theta-\theta^*)
\end{equation}
where $\theta^* \in \mathcal{S}$ is the unknown unconstrained minimizer, $J^* \in \mathbb{R}$, and $H \in \mathbb{R}^{n \times n}$ is a symmetric positive definite matrix (${H \succ 0}$).
\end{asmp}

\begin{asmp}[Barrier Function and Safe Set]
    \label{asmp:barrier-function}
    The barrier function $h : \mathbb{R}^n \to \mathbb{R}$ is twice differentiable ($h \in \mathcal{C}^2$), the safe set ${\mathcal{S} = \lbrace \theta \mid h(\theta) > 0 \rbrace}$ is non-empty, $\partial\mathcal{S}$ is compact, and $\nabla h(\theta) \neq 0$ for all $\theta\in\partial\mathcal{S}$. In addition, the sublevel sets of $h(\theta)$ are also compact.
\end{asmp}

The $\mathcal{C}^2$ regularity is required to ensure the existence and continuity of the Hessian $\nabla^2\hat{J}$, which appears in the local stability analysis.

\begin{thm}[Practical Safety via Forward Invariance]
\label{thm:safety}
    Consider system~\eqref{eq:dynamic-full} satisfying Assumptions~\ref{asmp:cost-function} and~\ref{asmp:barrier-function}. For any compact subset of initial conditions $\Theta_0 \subset \mathcal{S}$ away from the boundary of the safe set and a fixed $\mu>0$, there exists an $a^*$, dependent on $\Theta_0$ and the selected $\mu$, such that for all $a \in (0, a^*)$, there exists $\omega^*(\mu, a)$ such that for all $\omega\in(\omega^*, \infty)$, the safety constraint is strictly maintained for all $t \geq t_0$.
\end{thm}

Due to the singularity of the LBF term in \eqref{eq:modified-cost-function} as ${h(\theta) \to 0^{+}}$, the modified cost function $\hat{J}$ approaches infinity at the boundary of $\mathcal{S}$. Consequently, for any initial condition $\theta(t_0)$ belonging to a compact set $\Theta_0 \subset \mathcal{S}$ over which $\hat{J}$ is bounded, there exists a scalar $\underline{h} > 0$ which depends on $\Theta_0$ and $\mu$, such that $h(\theta(t)) \geq \underline{h} > 0$, for all $ t \geq t_0$ given appropriately selected $a$ and $\omega$. This implies that the safe set $\mathcal{S}$ is forward invariant for the closed-loop dynamics, ensuring that the system remains strictly safe and bounded away from the constraint boundary. The proof of Theorem~\ref{thm:safety} can be found in Section~\ref{sec:safety-proof}. While we can ensure practical safety with the correct choice of $a$ and $\omega$, we would like to seek a minimum, preferably the unconstrained minimum. As such, we give the local practical stability definitions, a Lemma for proving stability, and the stability theorem.

\begin{defn}[{modified from~\cite[Def~1]{ref:todorovski-2025}}]
    \label{def:lpuas}
    The system ${\dot{x} = f(t, x, \varepsilon)}$ with the vector of small parameters ${\varepsilon = [\varepsilon_1, \varepsilon_2, \dots, \varepsilon_m]^{\top}}$ is said to be locally practically uniformly asymptotically stable (LPUAS) to the origin if for every sufficiently small neighborhood  $\mathcal{B}\subset \mathbb{R}^n$ of the origin, there exist compact neighborhoods $\mathcal{W}, \mathcal{Q}$, and $\mathcal{V}$ of the origin with $\mathcal{Q} \subset \mathcal{B} \subset \mathcal{W}$ and $\mathcal{V} \subset \mathcal{W}$, and an $\varepsilon_1^* \in \mathbb{R}_{>0}$, such that for all $\varepsilon_1 \in (0, \varepsilon_1^*)$, there exists $\varepsilon_2^*$, such that for all $\varepsilon_2 \in (0, \varepsilon_2^*)$, $\dots$, there exists $\varepsilon_m^*$, such that for all $\varepsilon_m \in (0, \varepsilon_m^*)$, there exists $t_1 \in \mathbb{R}_{\geq 0}$ such that for all $t_0 \in \mathbb{R}$ the following holds:
    \begin{enumerate}
        \item (Boundedness) $x_0 \in \mathcal{V} \;\Rightarrow\; x(t) \in \mathcal{W} \quad \forall\, t \geq t_0$;
        \item (Stability) $x_0 \in \mathcal{Q} \;\Rightarrow\; x(t) \in \mathcal{B} \quad \forall\, t \geq t_0$;
        \item (Practical convergence) $x_0 \in \mathcal{V} \;\Rightarrow\; x(t) \in \mathcal{B}$ for all $t \geq t_1 + t_0$.
    \end{enumerate}
\end{defn}

\begin{lem}
\label{lem:proximity}
If $\theta^* \in \mathcal{S}$, then for sufficiently small $\mu$, there exists a $\hat{\theta}_{\mu}$ which is a minimizer of $\hat{J}$ and satisfies ${\Vert \hat{\theta}_\mu - \theta^* \Vert \in \mathcal{O}(\mu)}$.
\end{lem}
Lemma~\ref{lem:proximity} quantifies the distance between the critical points of $\hat{J}$ and the solutions of the original constrained optimization problem, establishing that the barrier introduces only an $\mathcal{O}(\mu)$ perturbation. The proof of Lemma~\ref{lem:proximity} is in Appendix~\ref{appendix:proof-of-proximity}.

\begin{thm}
\label{thm:stability}
    Consider the system~\eqref{eq:dynamic-full} satisfying Assumptions~\ref{asmp:cost-function} and~\ref{asmp:barrier-function}. If $\theta^* \in \mathcal{S}$, then the point $(\theta^*, J(\theta^*))$ is LPUAS with respect to the small parameter $(\mu, a, \omega^{-1})$.
\end{thm}
As one can see from Theorem~\ref{thm:stability}, we can alter the repulsiveness of the LBF through $\mu$ to get close to $\theta^*$, although the choice of $\mu$ affects the resulting necessary choices of $a$ and $\omega$. The proof of Theorem~\ref{thm:stability} can be found in Section~\ref{sec:stability-analysis}.

\section{Practical Safety Analysis}
\label{sec:safety-proof}
In this section, we provide the proof of Theorem~\ref{thm:safety}. The difficulty of the proof is that, unlike traditional practical stability results such as those in~\cite{ref:mcnamee-2025} or~\cite{ref:moreau-2000}, we cannot use an attractor to pull the trajectories further into the safe set. Instead, we need to use the repulsion of the LBF to prove the impermeability of a level set surrounding the unsafe sets. With appropriately selected dither amplitudes and rates, trajectories starting on the safe side of this impermeable set cannot traverse the level set to reach the unsafe set. However, the analysis proceeds in a similar manner to practical stability results with ESCs: 1) establish the average and model-based ESC systems; 2) demonstrate that the model-based ESC system strictly increases the value of $h$ near the boundary $h = 0$; and 3) show that through choices of $a$ and $\omega$, the model-free system is practically safe.

To aid our analysis, we use the following notation for the level, sublevel, and superlevel sets of $h$:
\begin{subequations}
\label{eq:level-set-notation}
    \begin{align}
    L_{c} &:= \left\{\theta \in \mathcal{S} \ \big|\  h(\theta) = c \right\} \\
    \underline{L}_c &:= \left\{\theta \in \overline{\mathcal{S}} \ \big|\  h(\theta) \leq c \right\} \\
    \overline{L}_c &:= \left\{\theta \in \mathcal{S} \ \big|\  h(\theta) \geq c \right\}
\end{align}
\end{subequations}
where $c > 0$. By Assumption~\ref{asmp:barrier-function}, all of the level and sublevel sets of $h(\theta)$ are compact.

\subsection{The Average and Model-Based Systems}
To aid in the analysis of the model-free ESC system, we first apply averaging theory to obtain the average system. Since all $\omega'_i$ are rational, there exists a common period $T$ for the signals $s(\tau)$ and $m(\tau, a)$. We denote the average system states by $\hat{\bar{\theta}}$ and their dynamics are given by
\begin{equation}
    \label{eq:avg-dynamics-approx}
    \dot{\hat{\bar{\theta}}} = -k \hat{\bar{g}}(\hat{\bar{\theta}}, a)
\end{equation}
where the averaged perturbation-based gradient estimate $\hat{\bar{g}}$ is defined as
\begin{equation}
    \hat{\bar{g}}(\hat{\bar{\theta}}, a) = \frac{1}{T} \int_0^{T} m(\tau) \hat{J}\left(\hat{\bar{\theta}} + a s(\tau)\right) d\tau.
\end{equation}
Since $\hat{J}$ is continuously differentiable by Assumptions~\ref{asmp:cost-function} and~\ref{asmp:barrier-function}, the average system becomes the model-based system
\begin{equation}
    \label{eq:avg-dynamics-approx-a-to-0}
    \dot{\vartheta} = -k \nabla \hat{J}(\vartheta)
\end{equation}
in the limit of $a \to 0$~\cite[Th~1]{ref:james-kavanaugh-2025}, with $\vartheta$ denoting the parameter estimate in the model-based system. Note that 
\begin{equation}
    \label{eq:c1-of-nabla-hat-j}
    \nabla \hat{J}(\theta) = H(\theta - \theta^*) - \mu \frac{\nabla h(\theta)}{h(\theta)} 
\end{equation}
is a standard gradient descent but now with a repulsive term to ensure strict safety.

\subsection{Safety Enforcement in the Model-Based System}
The Lie derivative of $h$ with the model-based system is
\begin{equation}
    \label{eq:h:lie-derivative}
    \dot{h}(\vartheta) = k \mu \frac{\Vert \nabla h(\vartheta) \Vert^2}{h(\vartheta)} - k \nabla J(\vartheta) \cdot \nabla h(\vartheta)
\end{equation}
To prove safety in the model-based system, we need to establish that $\dot{h} > 0$ whenever $h$ is sufficiently small. To do this, we need to establish some bounds on the terms in \eqref{eq:h:lie-derivative}. Note that as we are proving practical safety for the model-free system, we only need to consider the set of initial conditions $\Theta_0$ which have an $\underline{h} > 0$ such that
\begin{equation}
    h(\vartheta) \geq \underline{h}, \quad \forall\vartheta\in\Theta_0
\end{equation}

To start, Assumption~\ref{asmp:barrier-function} ensures that $\partial\mathcal{S}$ is compact and $\nabla h \neq 0$ on the boundary of the safe set, so there exists a constant $m > 0$ such that
\begin{equation}
    m = \min_{\theta\in\partial\mathcal{S}} \Vert \nabla h(\theta) \Vert
\end{equation}
Also, by $h\in\mathcal{C}^1$, there exists a $c_0 > 0$ such that
\begin{equation}
    \Vert \nabla h(\theta) \Vert \geq \frac{m}{2}, \quad \forall \theta\in\underline{L}_{c_0}
\end{equation}
and, without loss of generality, ${c_0 \leq \underline{h}}$. As $h\in\mathcal{C}^1$, we can establish the bound
\begin{equation}
    M_h := \max_{\theta \in \underline{L}_{c_0}} \Vert \nabla h(\theta)\Vert < \infty.
\end{equation}
Note that $\underline{L}_{c_0}$ is a compact set by Assumption~\ref{asmp:barrier-function}. Lastly, ${J(\theta)\in\mathcal{C}^1}$ by Assumption~\ref{asmp:cost-function}, so we can establish the bound
\begin{equation}
    M_J := \max_{\theta \in \underline{L}_{c_0}} \Vert H(\theta - \theta^*) \Vert < \infty.
\end{equation}
With all the necessary bounds found, we can bound \eqref{eq:h:lie-derivative} by
\begin{equation}
    \dot{h}(\vartheta)
    \geq - k\ M_h\ M_J + \frac{k \mu\,m^2}{4 h(\vartheta)}, \quad \forall \vartheta \in \underline{L}_{c_0}\cap\mathcal{S}
\end{equation}
using the Cauchy-Schwarz inequality. Therefore, we can glean that 
\begin{equation}
    \dot{h}(\vartheta)
        \geq \frac{k \mu\,m^2}{8 h(\vartheta)}, \quad \forall \vartheta\ \text{s.t.}\ h(\vartheta) \in \left(0, \frac{\mu\, m^2}{8 M_J M_h}\right].
\end{equation}
We will use $c_1 = (\mu\, m^2)/(8 M_J M_h)$ to represent this upper quantity and conclude that any trajectory of the model-based system with $\vartheta\in\Theta_0$ will have the strict safety margin of $c_1$. It is assumed without loss of generality that $c_1 \leq c_0$.

\subsection{Practical Safety Enforcement in the Average System}
Consider the Lie derivative of $h$ along the trajectories of the average system
\begin{align}
    \dot{h}(\hat{\bar{\theta}}, a) 
        &= \nabla h(\hat{\bar{\theta}}) \cdot \hat{\bar{g}}(\hat{\bar{\theta}}, a) \notag \\ 
        & \geq \begin{multlined}[t][0.7\columnwidth]
            -k \nabla h(\hat{\bar{\theta}}) \cdot\nabla \hat{J}(\hat{\bar{\theta}}) \\ - k \Vert \nabla h(\hat{\bar{\theta}}) \Vert\,\Vert \hat{\bar{g}}(\hat{\bar{\theta}}, a) - \nabla \hat{J}(\hat{\bar{\theta}}) \Vert
        \end{multlined}
\end{align}
Since the average system~\eqref{eq:avg-dynamics-approx} converges to the model-based system \eqref{eq:avg-dynamics-approx-a-to-0} as $a\to 0$, for any compact set $\hat{\bar{\theta}}$ where $\nabla\hat{J}$ is finite everywhere within, we can find an $a_1^*$ such that ${\Vert \hat{\bar{g}}(\hat{\bar{\theta}}, a) - \nabla \hat{J}(\hat{\bar{\theta}}) \Vert}$ is arbitrarily small for all $a\in(0,a_1^*)$. Note that the vector field depends on the value of $\mu$, so the convergence is dependent on its value, i.e., $a_1^*(\mu)$. For the analysis at hand, we will choose the set $\underline{L}_{c_1}\cap \overline{L}_{\frac{c_1}{4}}$ and an $a_1^*$ such that 
\begin{equation}
    \label{eq:h:average-system-diff-inequality}
    \dot{h}(\hat{\bar{\theta}}, a) \geq \frac{k \mu\,m^2}{16 h(\hat{\bar{\theta}})}, \quad \forall (\hat{\bar{\theta}}, a) \in (\underline{L}_{c_1}\cap \overline{L}_{\frac{c_1}{4}})\times(0,a_1^*)
\end{equation}
In doing so, we have demonstrated that the average system is practically safe with respect to the small parameter $a$. Furthermore, solving the differential inequality in \eqref{eq:h:average-system-diff-inequality} with an initial condition of $h(\hat{\bar{\theta}}_0) = \frac{c_1}{2}$ results in 
\begin{equation}
    h(\hat{\bar{\theta}}(t_1))^2 - \frac{c_1^2}{4} \geq \frac{k\mu\,m^2}{8}(t_1 - t_0)
\end{equation}
This means that after $(6 c_1^2)/(k\mu\,m^2)$ time has elapsed from the initial time $t_0$, the trajectory in the average system has an $h$ value of at least $c_1$ provided the initial $h$ value was at least $c_1 /2$.

\subsection{Practical Safety Enforcement in the Model-Free System}

Having established that the average system is practically safe with respect to the small parameter $a$, we will use this result to prove that the original model-free system is practically safe with respect to the small parameter vector $(a, \omega^{-1})$.

First, we need to establish the domain of $\hat{\theta}$ to which we will apply averaging results. Specifically, we will consider the domain $\underline{L}_{c_1}\cap \overline{L}_{\frac{c_1}{4}}$. The reason to consider this slightly larger domain than in the average system analysis is that we consider $\theta(t) = \hat{\theta}(t) + a s(\omega t)$ in our safety analysis and not just $\hat{\theta}$. Since $\Vert s(\tau) \Vert \leq R$ where $R\coloneqq \sqrt{\sum_{i=1}^{n}r_i^2}$, we want to find an $a_2^*$ such that $\theta$ does not enter the unsafe set for $a \in(0, a_2^*)$. To do this, we define two geometric quantities:
\begin{subequations}
\begin{align}
    \eta_1 &= \min_{x\in L_{c_1},\ y\in L_{\frac{c_1}{2}}} \Vert x - y \Vert \\
    \eta_2 &= \min_{x\in L_{\frac{c_1}{2}},\ y\in L_{\frac{c_1}{4}}} \Vert x - y \Vert 
\end{align}
\end{subequations}
The purpose of $\eta_1$ and $\eta_2$ is to establish how close the model-free trajectories must stay to the average system trajectories in our analysis domain and how large of an additive dither signal we are allowing. We choose ${a_2^* = \eta_2 /(2 R)}$ so that with $a^* = \min\lbrace a_1^*, a_2^*\rbrace$, all $a\in (0,a^*)$ ensures \eqref{eq:h:average-system-diff-inequality} holds and ${\Vert \theta(t) - \hat{\theta}(t) \Vert \leq \eta_2 /2}$.

For any fixed $a\in(0,a^*)$, we can use a standard averaging result, such as \cite[Th~10.4]{ref:khalil-2002}, when considering initial conditions $\hat{\theta}_0 \in L_{\frac{c_1}{2}}$ and trajectories within the domain $\underline{L}_{c_1}\cap \overline{L}_{\frac{c_1}{4}}$ to find $\omega^*$. This $\omega^*$ is a lower limit where for all $\omega > \omega^*$, trajectories of the average and model-free systems satisfy
\begin{equation}
    \Vert \hat{\theta}(t) - \hat{\bar{\theta}}(t) \Vert \leq \min\left\lbrace \eta_1, \frac{\eta_2}{2}\right\rbrace, \quad \forall\ t\in\left[t_0, t_0 + \frac{6 c_1^2}{k\mu\,m^2}\right]
\end{equation}
provided $\hat{\bar{\theta}}(t_0) = \hat{\theta}(t_0) \in L_{\frac{c_1}{2}}$. Let $\Delta = \min\left\lbrace \eta_1, \frac{\eta_2}{2}\right\rbrace$ and note that $\omega^*$ depends on the Lipschitz constant of the model-free system in the domain, which for our analysis depends on both $\mu$ and $a$. We express this dependency with $\omega^*(\mu, a)$.

We now have enough to establish the practical safety for the initial conditions $\hat{\theta}_0 \in \Theta_0$ with a proof by contradiction. Assume that, for a selection of $a\in(0,a^*(\mu))$ and ${\omega\in(\omega^*(\mu, a), \infty)}$, the system was not safe. Then there could be a trajectory with an initial condition $\hat{\theta}\in\Theta_0$ such that there is a time $t_{h=0}$, where $h(\theta(t_{h=0})) = 0$. As $\hat{\theta}_0\in\Theta_0$, we know that $h(\hat{\theta}_0) \geq \underline{h}\geq c_0 \geq c_1$. Therefore, there must be a ${t_1\in[t_0, t_{h=0}]}$ where $h(\hat{\theta}(t_1)) = \frac{c_1}{2}$, since neither the trajectories of $\theta(t)$ nor $\hat{\theta}(t)$ can reach $L_{0}$ without $\hat{\theta}(t)$ crossing through $L_{\frac{c_1}{2}}$ by both the geometric definition of $\eta_2$ and choice of $a < a^* \leq a_2^*$.

However, now the averaging results apply. Let $T = \frac{6 c_1^2}{k\mu\,m^2}$ so that $t_n = t_1 + (n-1)T$. If $\hat{\theta}(t_1) \in L_{\frac{c_1}{2}}$, then $\Vert \hat{\theta}(t) - \hat{\bar{\theta}}(t)\Vert \leq \Delta$ for all $t\in[t_1, t_2]$ with $\hat{\bar{\theta}}(t_1) = \hat{\theta}(t_1)$. By the triangular inequality
\begin{equation}
    \Vert \theta(t) - \hat{\bar{\theta}} (t) \Vert \leq \Vert \theta(t) - \hat{\theta} (t) \Vert + \Vert \hat{\theta}(t) - \hat{\bar{\theta}} (t) \Vert \leq \eta_2
\end{equation}
for $t\in[t_1,t_2]$. Since $h(\hat{\bar{\theta}}(t)) \geq \frac{c_1}{2}$ for all $t\in[t_1,t_2]$, $\theta(t)$ never crosses the boundary of $L_{\frac{c_1}{4}}$ in this time interval. By similar arguments, neither does $\hat{\theta}(t)$. Additionally, since ${\Vert \hat{\theta}(t_2) - \hat{\bar{\theta}}(t_2)\Vert \leq \eta_1}$ and $h(\hat{\bar{\theta}}(t_2)) \geq c_1$, the minimum value $h(\hat{\theta}(t_2))$ can take is $\frac{c_1}{2}$. Applying the average results inductively over the time intervals $[t_n, t_{n+1}]$ for $n=1,2,\ldots$ results in
\begin{subequations}
\begin{align}
    h(\hat{\theta}(t_1)) \geq \frac{c_1}{2} &\implies h(\hat{\theta}(t)) \geq \frac{c_1}{4}, \quad \forall t\in[t_1,\infty] \\
    h(\hat{\theta}(t_1)) \geq \frac{c_1}{2} &\implies h(\theta(t)) \geq \frac{c_1}{4}, \quad \forall t\in[t_1,\infty].
\end{align}
\end{subequations}
This contradicts the assumption of the existence of an unsafe trajectory, therefore, all trajectories with $\hat{\theta}_0\in\Theta_0$ must stay forever in the strict safe set for the selected $a$ and $\omega$. Thus, the result of Theorem~\ref{thm:safety} is proven, and the model-free system is practically safe with respect to the small parameter vector $(a, \omega^{-1})$. \hfill$\QEDopen$\hspace*{0.1cm}

\section{Stability Analysis}
\label{sec:stability-analysis}
In this section, we provide the proof for Theorem~\ref{thm:stability}. The proof relies on the characterization of equilibria, and applying the method of the averaging theorem~\cite[Th~1]{ref:mcnamee-2025}, and~\cite[Th~10.4]{ref:khalil-2002} to establish the model-free system stability.

\subsection{Characterization of Equilibria}
Unlike the standard convex setting, the modified cost function $\hat{J}(\theta)$ is not guaranteed to be globally convex, which means multiple equilibria may exist. The equilibria of the model-based system~\eqref{eq:avg-dynamics-approx-a-to-0} correspond to the critical points of the modified cost function, defined as
\begin{equation}
    \hat{\mathcal{E}}_\mu = \{ \theta \in \mathcal{S} \mid \nabla \hat{J}(\theta) = 0 \}.
\end{equation}
The gradient condition \eqref{eq:c1-of-nabla-hat-j} represents a force balance between the attractive force of the quadratic cost $H(\theta - \theta^*)$ and the repulsive force of the barrier $\mu \frac{\nabla h(\theta)}{h(\theta)}$. As established in Lemma~\ref{lem:proximity}, the set $\hat{\mathcal{E}}_\mu$ is an $\mathcal{O}(\mu)$ perturbation of the original unconstrained minimizers $\mathcal{E}^*$.

\subsection{Convergence Analysis}
We analyze the stability of the system by proceeding through three steps. We first establish the convergence of the model-based system, and then quantify the exponential rate of the model-based system. After that, we transfer these properties to the averaged system at finite $a$, and then apply finite-time averaging to the original time-varying system. Finally, we combine the bounds via the triangle inequality to obtain the original time-varying system properties.

\paragraph{Convergence of the model-based system} Consider a candidate Lyapunov function $V = \hat{J}(\vartheta) - \hat{J}(\hat{\theta}_{\mu})$ for the model-based system~\eqref{eq:avg-dynamics-approx-a-to-0}. Then, we obtain the time derivative along trajectories
\begin{equation}
    \dot{V} = \nabla \hat{J}(\vartheta)^{\top} \dot{\vartheta} = -k \Vert \nabla \hat{J}(\vartheta) \Vert^2 \leq 0.
\end{equation}
with equality if and only if $\vartheta = \hat{\theta}_{\mu}$. According to Lemma~\ref{lem:proximity}, every trajectory initialized in $\Theta_0$ converges asymptotically to $\hat{\theta}_\mu$.

To quantify the convergence rate, we examine the Hessian of $\hat{J}$ at $\hat{\theta}_\mu$. Since $\hat{\theta}_\mu \to \theta^*$ as ${\mu \to 0}$, and ${\nabla^2 \hat{J} \to H}$ as $\mu \to 0$ in a neighborhood of $\theta^*$, there exists $\mu_1^* > 0$ such that for all $\mu \in (0, \mu_1^*)$, we obtain ${\nabla^2 \hat{J}(\hat{\theta}_\mu) \succeq \frac{\lambda_{\min}(H)}{2} I}$. Consequently, the linearization of the model-based system~\eqref{eq:avg-dynamics-approx-a-to-0} at $\hat{\theta}_\mu$ yields a locally exponentially stable equilibrium with a convergence rate of at least $\frac{k \lambda_{\min}(H)}{2}$.

\paragraph{Equilibrium and stability of the averaged system at finite $a$} The averaged system~\eqref{eq:avg-dynamics-approx} involves the averaged gradient $\hat{\bar{g}}(\hat{\bar{\theta}}, a)$, which satisfies $\hat{\bar{g}}(\cdot, a) \to \nabla \hat{J}(\cdot)$ as $a \to 0$. According to Assumptions~\ref{asmp:cost-function} and~\ref{asmp:barrier-function}, we know that $\hat{J} \in \mathcal{C}^2$ on $\mathcal{S}$, and the averaged gradient $\hat{\bar{g}}(\cdot, a)$ is a continuously differentiable function in the first argument, whose Jacobian converges to $\nabla^2 \hat{J}$ as $a \to 0$. According to~\cite[Th~1]{ref:mcnamee-2025}, for sufficiently small $a > 0$, the averaged system~\eqref{eq:avg-dynamics-approx} admits a unique equilibrium $\hat{\theta}_{\mu,a}$ in a neighborhood of $\hat{\theta}_\mu$ and satisfies $\Vert\hat{\theta}_{\mu,a} - \hat{\theta}_\mu\Vert = \mathcal{O}(a)$ for \emph{fixed} $\mu$. Moreover, the Jacobian of the averaged system at $\hat{\theta}_{\mu,a}$ is a continuous perturbation of $\nabla^2 \hat{J}(\hat{\theta}_\mu)$, so for $\mu \in (0, \mu_1^*)$ and the dither amplitude $a$ sufficiently small, the equilibrium $\hat{\theta}_{\mu,a}$ is locally exponentially stable with a convergence rate of at least $\frac{k \lambda_{\min}(H)}{4}$.

\paragraph{Stability of the original system} Since the averaged system~\eqref{eq:avg-dynamics-approx} possesses the locally exponentially stable equilibrium $\hat{\theta}_{\mu,a}$ within the compact sublevel set $\Theta_0$, and the vector field and its derivatives are uniformly bounded on $\Theta_0$, the averaging theorem~\cite[Th~10.4]{ref:khalil-2002} guarantees that for $\omega$ sufficiently large, the trajectories of the original time-varying system~\eqref{eq:dynamic-full} converge to an $\mathcal{O}(\omega^{-1})$ neighborhood of $\hat{\theta}_{\mu,a}$. Combining all the bounds via the triangle inequality, we obtain
\begin{equation}
    \Vert\hat{\theta}(t) - \theta^*\Vert \leq  \Vert\hat{\theta}_\mu - \theta^*\Vert+ \Vert\hat{\theta}_{\mu,a} - \hat{\theta}_\mu\Vert + \Vert\hat{\theta}(t) - \hat{\theta}_{\mu,a}\Vert 
\end{equation}
for all $t \geq t_0$. Therefore, given a desired $\Delta$, we choose a $\mu^*$ such that the first term is less than $\Delta/3$ for any $\mu \in (0, \mu^*)$. For the second term, given a fixed $\mu$, we can find an $a^*(\mu)$ such that for any $a$ less than $a^*$, the second term is less than $\Delta/3$. The last term can also be made arbitrarily small in a manner similar to the first two terms. For fixed parameters $a$ and $\mu$, we can find an $\omega^*(\mu, a)$ such that, for all $\omega > \omega^*$, the last term is less than $\Delta /3$. Thus, by sequential selections of $\mu$, $a$, and then $\omega$, we can ensure that the model-free system~\eqref{eq:dynamic-full} converges to an arbitrarily small $\Delta$ neighborhood of $\theta^*$, completing the proof of Theorem~\ref{thm:stability}. \hfill $\QEDopen$\hspace*{0.1cm}

\section{Simulation Results}
This section compares our LBF approach against both a standard, unconstrained, gradient-based ESC and the CBF-based method developed in \cite{ref:williams-2022}. 

\subsection{1D Case Comparison}
\begin{figure}[t]
    \vspace*{0.3cm}
    \centering
    \includegraphics[width=0.8\columnwidth]{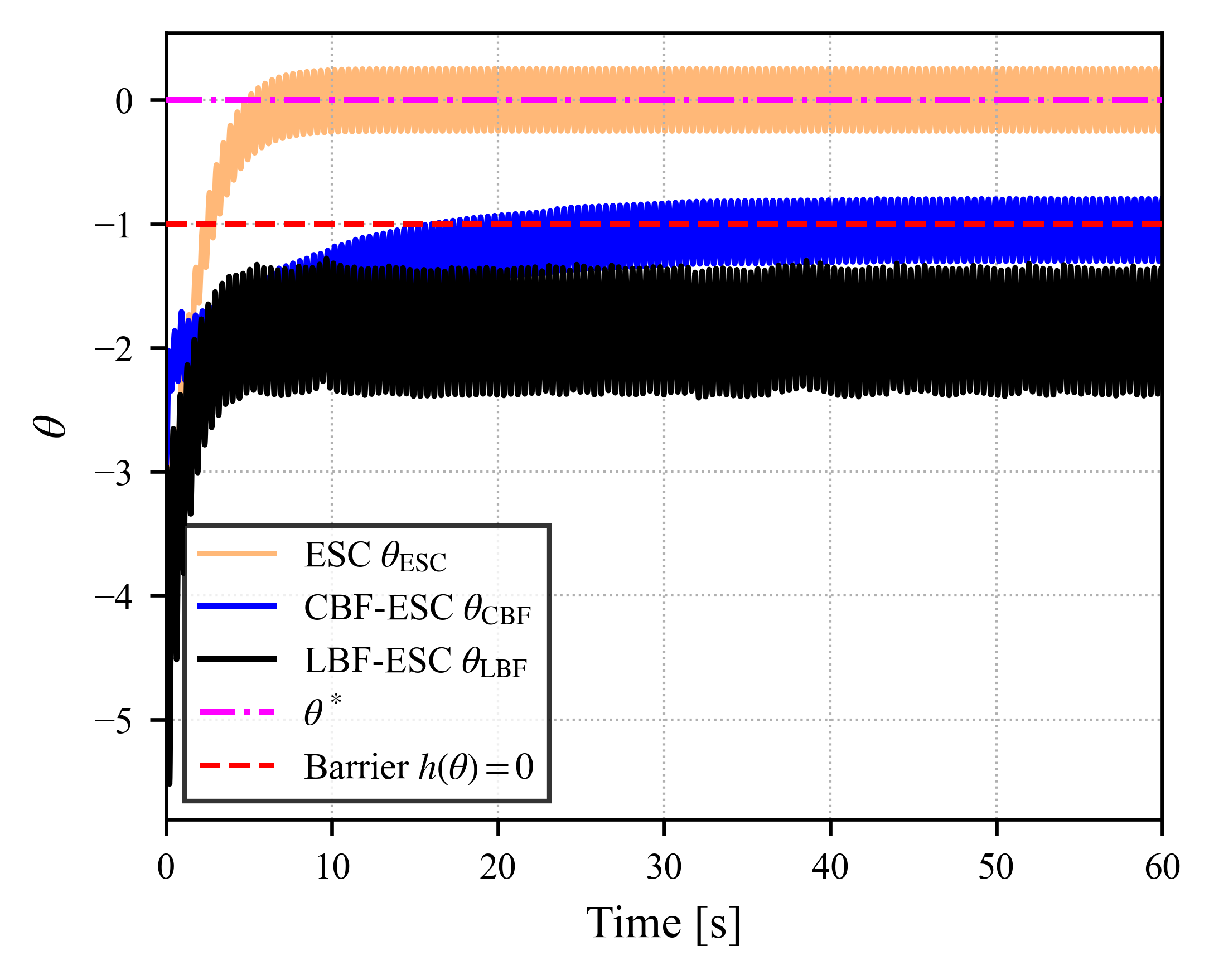}
    \caption{Simulation in the 1D case. The initial position of the seeker is ${\theta_{\text{ESC}} = \theta_{\text{LBF}} = \theta_{\text{CBF}} = -3}$, and the unconstrained optimum is at $\theta^* = 0$. The controller parameters are set as follows: $a = 0.25$, $k = 0.2$, $\omega = 15$, and $\mu = 3$ for the LBF-ESC; and $a = 0.25$, $k = 0.3$, $\omega = 15$, $\omega_h=\omega_\ell=4.5$, $c = 0.1$, and $\delta = 0.001$ for the CBF-ESC.}
    \label{fig:1d}
\end{figure}

Following \cite{ref:williams-2022}, we define the safety function as $ {h(\theta) = -\theta - 1}$, where the safe set is $\{\theta \mid \theta < -1 \}$, and the cost function as $ J(\theta) = \theta^2 $. The simulation results of an unconstrained gradient-based ESC (denoted ESC), the ESC using CBF methods from \cite{ref:williams-2022} (denoted CBF-ESC), and our ESC using the LBF method (denoted LBF-ESC) are shown in Fig.~\ref{fig:1d}. While the unconstrained ESC converges to the global optimum $\theta^* = 0$, it violates the safety constraint. The CBF–ESC performs better, converging to the constrained minimizer, but there is also a violation of the safety constraint near the boundary of $\theta = -1$. In contrast, the proposed LBF–ESC converges near the constrained optimum without violating the safety constraint. The practical convergence of the LBF-ESC can be made closer to the constrained minimizer by choosing a smaller $\mu$, but the emphasis of the LBF-ESC is the assurance of strict safety, something that the unconstrained ESC and the CBF-ESC do not guarantee.

\subsection{2D Case Comparison}
\subsubsection{Complex Nonlinear Constraints}
\label{sec:simulations:2d:corridor}
\begin{figure}
    \centering
    \includegraphics[width=0.8\columnwidth]{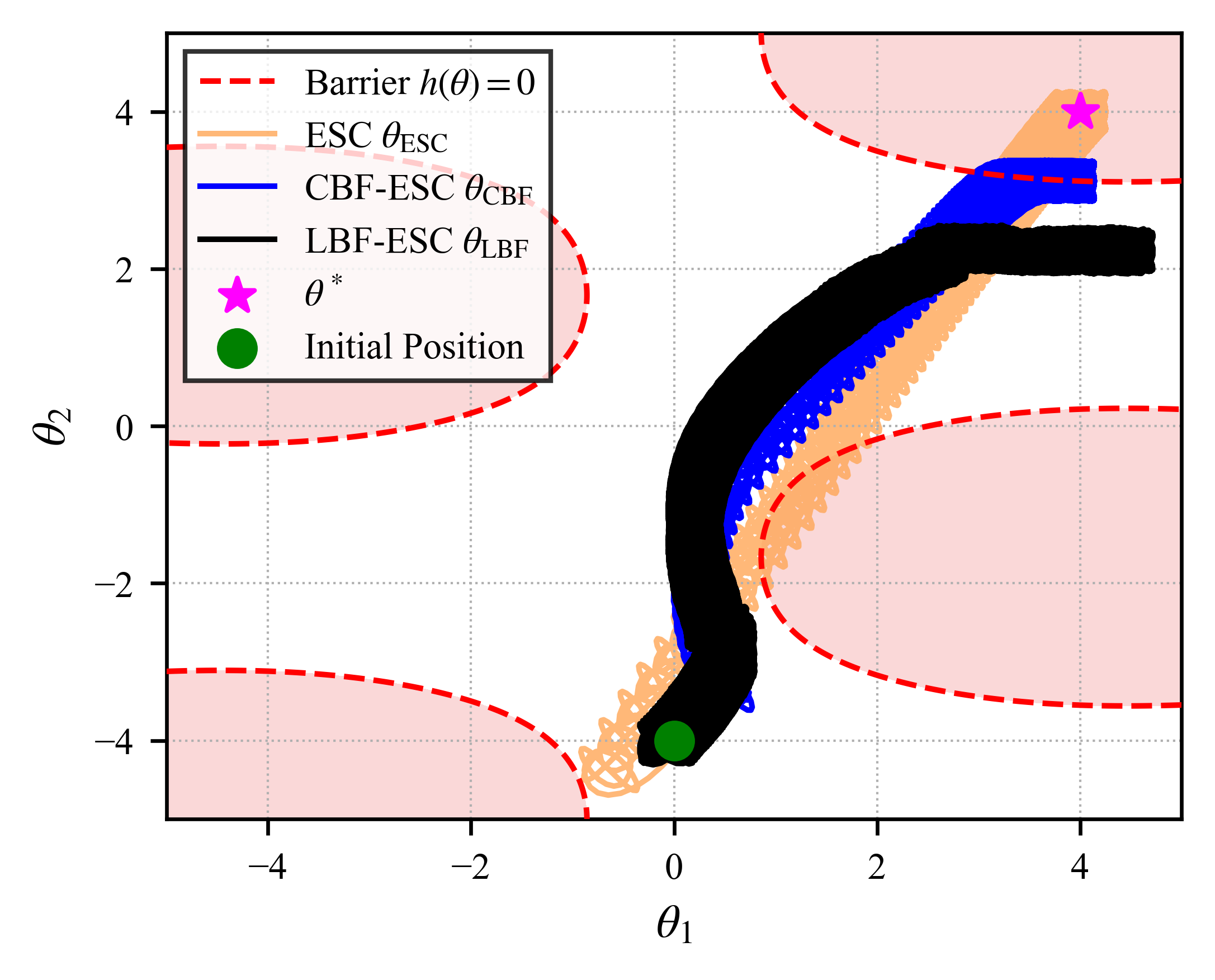}
    \caption{Simulation of a 2D point mass case with island-shaped obstacle avoidance. The initial position of the seeker is $(\theta_1, \theta_2) = (0, -4)$, and the unconstrained optimum is at $\theta^* = (4, 4)$. The controller parameters are set as follows: $a = 0.25$, $k = 0.01$, $\omega_1 = 75$, $\omega_2 = 100$, and $ \mu = 6$ for the LBF-ESC; and $a = 0.25$, $k = 0.1$, $\omega_1 = 75$, $\omega_2 =  100$, $\omega_h=\omega_\ell=30$, $c = 0.5$, and $\delta = 0.001$ for the CBF-ESC.}
    \label{fig:2d-cos}
\end{figure}

In addition to the 1D scenario, we also adapt a 2D scenario from \cite{ref:williams-2022} where the safety filter is defined by the nonlinear function ${h(\theta) = \cos(0.2\pi\theta_1)\sin(0.3\pi\theta_2)} $, and the cost function is $ J(\theta) = (\theta_1 - 4)^2 + (\theta_2 - 4)^2 $. Figure~\ref{fig:2d-cos} shows the trajectories of the three different methods for this scenario, all using relative dither amplitudes of $r_1 = r_2 = 1$. Similar to the CBF-ESC, the LBF-ESC successfully navigates the complex non-convex safe set, keeping the agent within the safe region while converging to a neighborhood of the constrained optimum permitted by the barrier function. However, as in the 1D scenario, the LBF-ESC is the only ESC method that maintains strict safety throughout its trajectory.

\subsubsection{Corridor Navigation}
\begin{figure}[t]
    \vspace*{0.3cm}
    \centering
    \includegraphics[width=0.8\columnwidth]{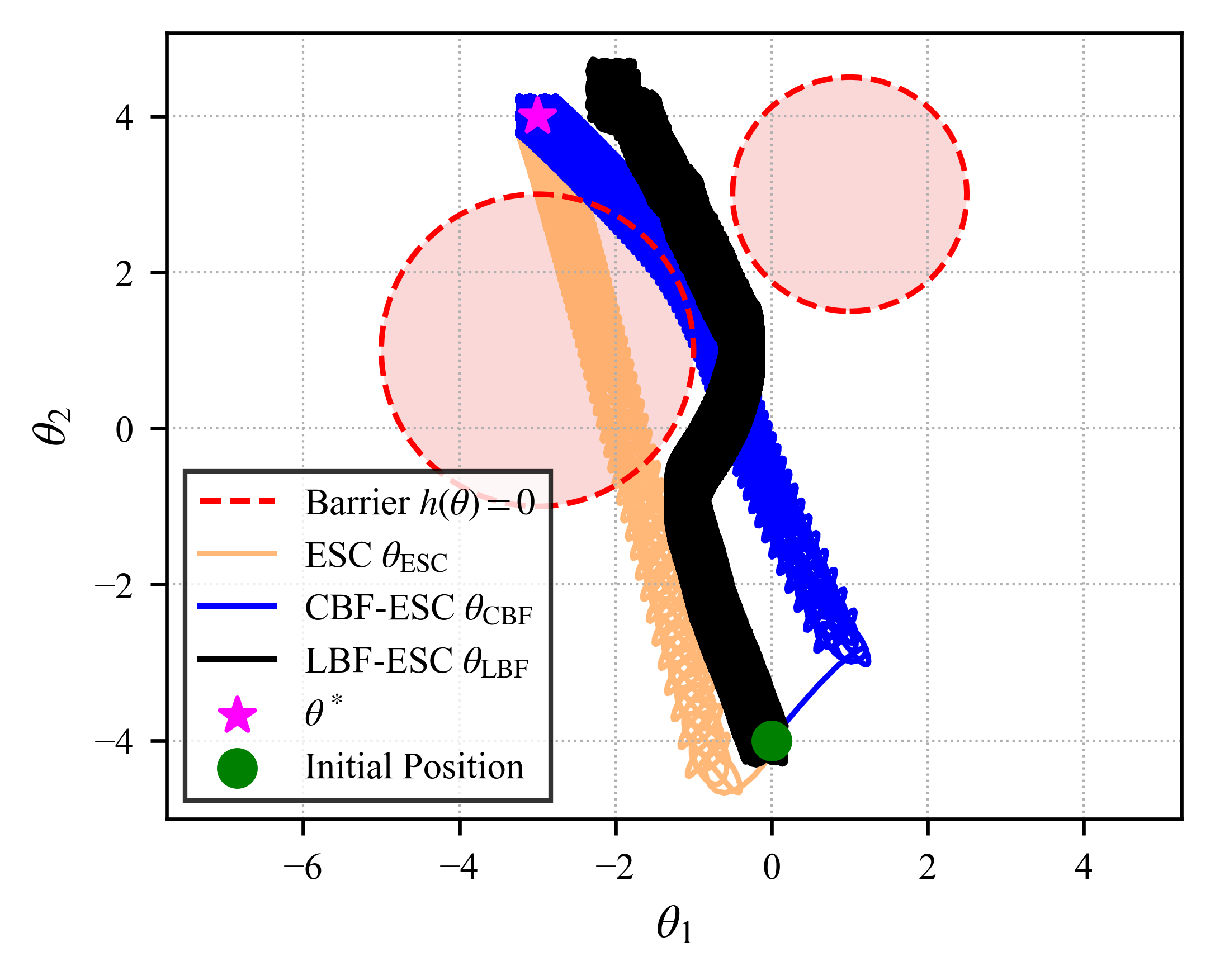}
    \caption{Simulation of a 2D point mass case with passage navigation. The obstacles are located at $(-3, 1)$ and $(1, 3)$ with radii equal to $2$ and $1.5$, respectively. The initial position of the seeker is $(\theta_1, \theta_2) = (0, -4)$, and the unconstrained optimum is at $\theta^* = (-3, 4)$. The controller parameters are set as follows: $a = 0.25$, $k = 0.01$, $\omega_1 = 75$, $\omega_2 =  100$, and $\mu = 6$ for the LBF-ESC; and $a = 0.25$, $k = 0.1$, $\omega_1 = 75$, $\omega_2 = 100$, $\omega_h=\omega_\ell=30$, $c = 1$, and $\delta = 0.001$ for the CBF-ESC.}
    \label{fig:2d-circle}
\end{figure}

In real-world source-seeking scenarios, one difficulty is safely navigating the narrow corridors that form between two (or more) obstacles to reach the source. This is a more challenging task than the one originally presented in \cite{ref:williams-2022}, where the obstacles are spaced far apart relative to the dither amplitude. To demonstrate this more challenging navigation scenario, we utilize an environment with two circular obstacles, where the seeker is required to traverse between them to converge to the source in the shortest time. The barrier function for the two circular obstacles is
\begin{equation}
    h(\theta) = \min_{i=1,2}\left\lbrace\Vert\theta - \theta_{\text{center},i}\Vert - r_i\right\rbrace
\end{equation}
where $\theta_{\text{center},i}$ is the center of the $i$th circular obstacle and $r_i$ is its radius. Simulations with a cost function of ${J(\theta) = (\theta_1 +3)^2 + (\theta_2 - 4)^2}$, the same dither signal in Section~\ref{sec:simulations:2d:corridor}, and a width corridor of slightly less than $1$ are shown in Fig.~\ref{fig:2d-circle}. The CBF-ESC shows the same constraint violation as in the previously shown 1D simulation result, while the LBF-ESC maintains strict safety throughout the corridor navigation scenario. The CBF-ESC practically converges to the unconstrained minimizer, but the LBF-ESC has an offset from the unconstrained minimizer. This offset is an inherent property of LBF methods, representing the equilibrium between the attraction of the cost function and the repulsion of the safety barrier. This trade-off allows for tunable conservatism, with trajectories of the LBF-ESC system settling closer to the unsafe set with a smaller $\mu$ and farther from the unsafe set with a larger $\mu$.

\section{Conclusion}
This paper presents a methodology for integrating LBFs with ESCs, achieving model-free optimization with provable practical safety guarantees. By incorporating the LBF into the objective function, the controller inherently balances performance with safety constraints. Importantly, the proof of practical safety in this work relied on the \emph{repulsion} of the barrier rather than the \emph{attraction} of the minimum in the constrained optimization problem. This allowed us to use weaker assumptions than in the previous literature, extending practical safety results from problems only considering concave $h$ (cf.~\cite{ref:labar-2019,ref:williams-2025,ref:williams-2026}) to those considering potentially non-concave $h$. We used simulations to illustrate the results of our methodology against existing CBF-based methodologies in the previous literature. The methodology presented in this work was able to practically converge to the sources, similar to results found in previous literature, but our LBF-based approach ensured the required strict safety constraint.




\appendices
\section{Proof of Lemma~\ref{lem:proximity}}
\label{appendix:proof-of-proximity}
The proof considers the case in which the unconstrained minimizer $\theta^*$ is in the open set $\mathcal{S}$. Since the critical point $\hat{\theta}_\mu$ satisfies $\nabla\hat{J}(\hat{\theta}_\mu) = 0$, we obtain ${H(\hat{\theta}_\mu - \theta^*) = \mu \frac{\nabla h(\hat{\theta}_\mu)}{h(\hat{\theta}_\mu)}}$. We define the mapping $F(\theta, \mu) = H(\theta - \theta^*) - \mu \frac{\nabla h(\theta)}{h(\theta)}$. It follows directly that $F(\theta^*, 0) = 0$. The Jacobian of $F$ with respect to $\theta$ at $(\theta^*, 0)$ is $\frac{\partial F}{\partial \theta} = H$. Because $H \succ 0$, the Jacobian is invertible. The Implicit Function Theorem~\cite[p651]{ref:khalil-2002} shows that there exists a smooth curve $\mu \mapsto \hat{\theta}_\mu$, which is defined for $\mu$ in a neighborhood of $0$. The curve satisfies the constraints $F(\hat{\theta}_\mu, \mu) = 0$ and $\hat{\theta}_0 = \theta^*$.
 
To show the closeness of $\hat{\theta}_{\mu}$ to $\theta^*$, we differentiate $F(\hat{\theta}_\mu, \mu) = 0$ with respect to $\mu$ at $\mu = 0$, and obtain the relation $ H \frac{d\hat{\theta}_\mu}{d\mu}\bigg|_{\mu=0} = \frac{\nabla h(\theta^*)}{h(\theta^*)}$. By a Taylor series expansion, ${\hat{\theta}_\mu = \theta^* + \mu H^{-1} \frac{\nabla h(\theta^*)}{h(\theta^*)} + \mathcal{O}(\mu^2)}$, which means ${\Vert \hat{\theta}_\mu - \theta^*\Vert = \mathcal{O}(\mu)}$. We note that the displacement $\hat{\theta}_\mu - \theta^*$ is approximately $\mu H^{-1}\frac{\nabla h(\theta^*)}{h(\theta^*)}$. Since $H \succ 0$, the inner product $\nabla h(\theta^*)^\top (\hat{\theta}_\mu - \theta^*)\approx \frac{\mu}{h(\theta^*)}\nabla h(\theta^*)^\top H^{-1}\nabla h(\theta^*)$ is strictly positive. This implies that the displacement has a positive projection onto the gradient $\nabla h(\theta^*)$, effectively pointing into the interior of $\mathcal{S}$. Consequently, the modified equilibrium $\hat{\theta}_\mu$ satisfies $h(\hat{\theta}_\mu) > h(\theta^*)$, for $\mu > 0$ sufficiently small. That is, the barrier repulsion pushes the equilibrium strictly deeper into the safe set compared to the unconstrained minimizer, providing an additional safety margin that grows with $\mu$.

To show that $\hat{\theta}_{\mu}$ is a local minimum, we examine the second derivative of $\hat{J}$ in the neighborhood of $\hat{\theta}_{\mu}$. By Assumptions~\ref{asmp:cost-function} and~\ref{asmp:barrier-function}, the second derivative exists
\begin{equation}
    \nabla^2 \hat{J}(\theta) = H - \frac{\mu}{h(\theta)} \nabla^2 h(\theta) 
    + \frac{\mu}{h(\theta)^2} \nabla h(\theta) \nabla h(\theta)^{\top}
\end{equation}
within $\mathcal{S}$. Within a sufficiently small neighborhood of $\hat{\theta}_{\mu}$, the quantities $h$ and $\nabla^2 h$ are bounded, thus, it is always possible to find a $\mu^*$ such that $\nabla^2\hat{J}\succ \frac{1}{2}\lambda_{\min}(H)$ for all $\mu\in(0,\mu^*)$. Therefore, we can select $\mu$ sufficiently small so that the modified cost function is strongly convex and, consequently, $\hat{\theta}_\mu$ must be a local minimum. \hfill $\QEDopen$



\bibliographystyle{IEEEtranS}
\bibliography{citation}


\end{document}